\documentclass[12pt]{iopart}

\usepackage{iopams}
\usepackage {graphicx}

\begin{document}
\title{Pinning Enhancement by Heterovalent Substitution in Y$_{1-x}$RE$_{x}$Ba$_{2}$Cu$_{3}$O$_{7-\delta}$}

\author{M I Petrov, Yu S Gokhfeld, D A Balaev, S I Popkov, A A
Dubrovskiy, D M Gokhfeld, K A Shaykhutdinov}

\address{L.V. Kirensky Institute of Physics SD RAS, 660036 Krasnoyarsk,
Russia}

\ead{smp@iph.krasn.ru}

\begin{abstract}
The intragrain pinning in high-$T_c$ superconductor compounds
Y$_{1-x}$RE$_{x}$Ba$_{2}$Cu$_{3}$O$_{7-\delta }$ with low
concentration of RE (La, Ce, Pr) was investigated. Magnetic and
transport measurements reveal that the pinning is maximal for the
concentration of heterovalent RE such that the average distance
between the impurity ions in the plane of rare-earth elements close
to the diameter of Abrikosov vortices in YBCO.
\end{abstract}


\section{Introduction}

Improvement of critical current of YBCO materials is attained by a
creation of additional defects acting as pinning centers.
Irradiation, incorporation of nanoparticles and doping
\cite{pauli,skakle,hara1,hara2,xing} are the main ways to increase
the pinning. In last case the partial substitution of rare earth
elements (RE) for Y is favorable \cite{hara1,macma,kell,barnes,huht}
that results in a local distortions of crystal structure and
electron density. Y$_{1-x}$RE$_{x}$Ba$_{2}$Cu$_{3}$O$_{7-\delta }$
was earlier investigated in most cases with the concentration $x$
about a few tens of atomic percents. It is found that the pinning
depends on valence and size of RE ions. The doping ions with valence
3+ do not change superconducting properties of YBCO greatly unlike
ions with larger valence \cite{taras}. Increasing of pinning in
Y$_{1-x}$RE$_{x}$Ba$_{2}$Cu$_{3}$O$_{7-\delta }$ films with small
concentrations of RE ($x < $ 0.1) was observed in article
\cite{barnes}. Authors of \cite{barnes} were interested mainly by a
relative influence of different doping elements to the flux pinning.
They found the pinning enhancement by a minute doping with different
RE especially Tb and Nd. The dependence of critical current density
in polycrystalline Y$_{1-x}$Pr$_{x}$Ba$_{2}$Cu$_{3}$O$_{7}$ on $x$
was investigated in article \cite{huht}. The maximal critical
current density was reached for $x$ = 0.08.

To study the concentration dependence of pinning we suggest to
choose $x$ connecting with parameters of the crystal structure. The
concentration of RE can be correlated with the average distance
between impurity ions in the rare-earth plane, $D$. Atoms of Y
arrange the planes in YBCO such that the connection between $x$ and
$D$ is given by $x = a^{2}/D^{2}$, where $a$ is the lattice constant
in the rare-earth plane. It can be written as $x = a^{2}/(na)^{2} =
1/n^{2}$, where $n = D/a$. Thus one can choose $x$ to obtain integer
$n$ = 2, 3, 4, 5, 6, 7, 8, 9, 10, $\infty $, i.e. $D$ is divisible
by $a$. The compound with $n = \infty $ corresponds to classical
YBa$_{2}$Cu$_{3}$O$_{7-\delta }$. Such choice of $x$ reveals the
pinning dependence on the average distance between the pinning
defects.

To prove an influence of the heterovalent substitution on the
pinning we investigate YBCO doped by Ce and Pr. Ions Ce and Pr have
valence equal 4+ and 3$\div$4+ correspondingly and should strongly
modify superconductivity near their location. Ions La have the same
valence as Y but the larger radius. The doping by La is chosen to
examine the influence of a lattice distortion without the distortion
of electron density.

Earlier we published the first results for
Y$_{1-x}$Ce$_{x}$Ba$_{2}$Cu$_{3}$O$_{7 - \delta }$ with the
above-stated concentrations $x$ \cite{PhC07,FTT07}. Here we report
about comparative study of the magnetic and transport
characteristics of Y$_{1-x}$Pr$_{x}$Ba$_{2}$Cu$_{3}$O$_{7 - \delta
}$, Y$_{1-x}$La$_{x}$Ba$_{2}$Cu$_{3}$O$_{7 - \delta }$ and
Y$_{1-x}$Ce$_{x}$Ba$_{2}$Cu$_{3}$O$_{7 - \delta }$. Measurements of
magnetization and resistance were carried out. It allows comparing
the intragrain pinning in the compounds with different $x$.

\section{Experiment}

Three series of compositions
Y$_{1-x}$RE$_{x}$Ba$_{2}$Cu$_{3}$O$_{7-\delta }$ were prepared for
RE = Ce, Pr, La. The series (Y$_{1-x}$Ce$_{x}$Ba$_{2}$Cu$_{3}$O$_{7
- \delta }$, Y$_{1-x}$Pr$_{x}$Ba$_{2}$Cu$_{3}$O$_{7 - \delta }$ and
Y$_{1-x}$La$_{x}$Ba$_{2}$Cu$_{3}$O$_{7 - \delta }$) were synthesized
separately by the standard solid-phase technique. The starting
reactants were Y$_2$O$_3$, CeO$_2$, Pr$_6$O$_{11}$, La$_2$O$_3$,
CuO, BaCO$_3$. The corresponding amounts of reagents were mixed
thoroughly in an agate mortar, pelletized, and annealed at 930
$^0$C. The synthesis, including seven intermediate crushings and
pressings, lasted 160 h. Long procedures favor ordering of RE
elements and cerium substitution in yttrium positions. The synthesis
completed, the samples were annealed at a temperature of 300 $^0$C
for 3 h and cooled slowly in the furnace to room temperature to
reach oxygen saturation.

Each series contain 10 samples with $x$ = 0.25, 0.11, 0.0625, 0.04,
0.0278, 0.0204, 0.0156, 0.0123, 0.01 and 0 which were synthesized
simultaneously at the same conditions. The chosen concentrations of
RE correspond to $n$ = 2, 3, 4, 5, 6, 7, 8, 9, 10 and $\infty $. In
such a way the composition with $x$ = 0 (YBa$_{2}$Cu$_{3}$O$_{7 -
\delta })$ was synthesized for each series.

The temperature dependence of resistance $R(T)$ was measured by the
standard four-probe technique with the bias current 10 mA. Samples
have rectangular cross section (2 mm $\times $1 mm), the distance
between the potential contacts being 2 mm.

Magnetic characteristics were measured by a vibrating sample
magnetometer. The samples were cut out in the shape of cylinders
with height $ \approx $ 5 mm and diameter $ \approx $ 0.5 mm. The
temperature dependence of magnetization is measured at the samples
cooled without external field ($M_{ZFC})$ and in magnetic field 100
Oe ($M_{FC})$. The measurements were carried out during heating with
speed 0.8 K/min from 77.4 K to 100 K for
Y$_{1-x}$Ce$_{x}$Ba$_{2}$Cu$_{3}$O$_{7 - \delta }$ and
Y$_{1-x}$La$_{x}$Ba$_{2}$Cu$_{3}$O$_{7 - \delta }$ and from 55 K to
100 K for Y$_{1-x}$Pr$_{x}$Ba$_{2}$Cu$_{3}$O$_{7 - \delta }$. The
magnetic field $H$ = 100 Oe was applied parallel to the cylinder
axis. The temperature of intragrain superconducting transition
$T_{c}$ was determined from $M_{ZFC}(T)$ (criterion is d$M$/d$T$ =
0). The temperature of disappearance of resistance  $T_{c0}$ was
determined from $R(T)$ (criterion is voltage drop on sample = 1
$\mu$V cm).

\section{Results and Discussion}

The x-ray diffractions patterns show that most of samples are
single-phase and have YBCO structure.  There are small distortions
of crystal lattice for Y$_{1-x}$La$_{x}$Ba$_{2}$Cu$_{3}$O$_{7 -
\delta }$ with high concentrations of La ($x $= 0.11, $x $= 0.25).
For Y$_{1-x}$Ce$_{x}$Ba$_{2}$Cu$_{3}$O$_{7 - \delta }$ the phase
BaCeO$_{3}$ precipitates when $x >$ 0.024 (see details in
\cite{FTT07}) because Ce dissolves in YBCO at low concentrations
only.

\begin{figure}[htbp]
\centerline{\includegraphics[width=6.16in,height=4.84in]{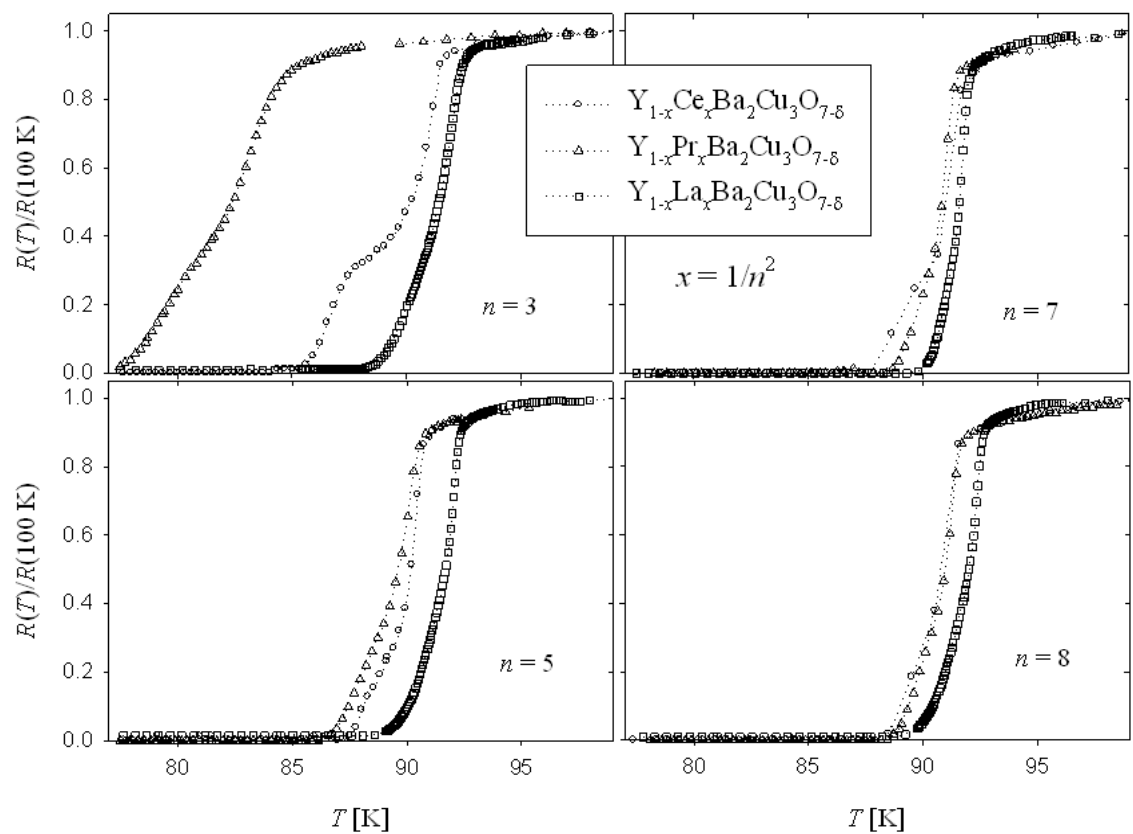}}
\caption{Temperature dependence of normalized resistance
$R(T)$/$R$(100 K) of samples
Y$_{1-x}$RE$_{x}$Ba$_{2}$Cu$_{3}$O$_{7-\delta}$.}
\label{fig1}
\end{figure}

\Fref{fig1} displays the temperature evolution of the resistance
$R(T)$ normalized to $R$(100 K) for a few samples. The measured
$R(T)$ dependences are typical for polycrystalline superconductors
and exhibit a sharp drop of the resistance at $T_{c}$ and a smooth
part till $T_{c0}$ reflecting the transition of Josephson media
formed by the intergrain boundaries. Above $T_{c}$ most of
dependences $R(T)$ are metallic like. The exception is
Y$_{0.75}$La$_{0.25}$Ba$_{2}$Cu$_{3}$O$_{7 - \delta }$ ($n$ = 2)
having a quasi-semiconductor like $R(T)$ above $T_{c}$. It is
probably because La occupies partially Ba sites for high $x$
\cite{gantis}. The dependences $R(T)$ demonstrate that Pr depresses
superconductivity stronger than Ce or La. The remarkable tail from
$T_{c}$ to $T_{c0}$ on $R(T)$ of
Y$_{1-x}$Ce$_{x}$Ba$_{2}$Cu$_{3}$O$_{7 - \delta }$ for $n$ = 3 shows
that the intergrain boundaries thickness is increased due to a
precipitating of nonsuperconducting phase (see the article
\cite{sust04} concerning the composites YBCO + nonsuperconducting
compounds).

\begin{figure}[htbp]
\centerline{\includegraphics[width=5.81in,height=4.69in]{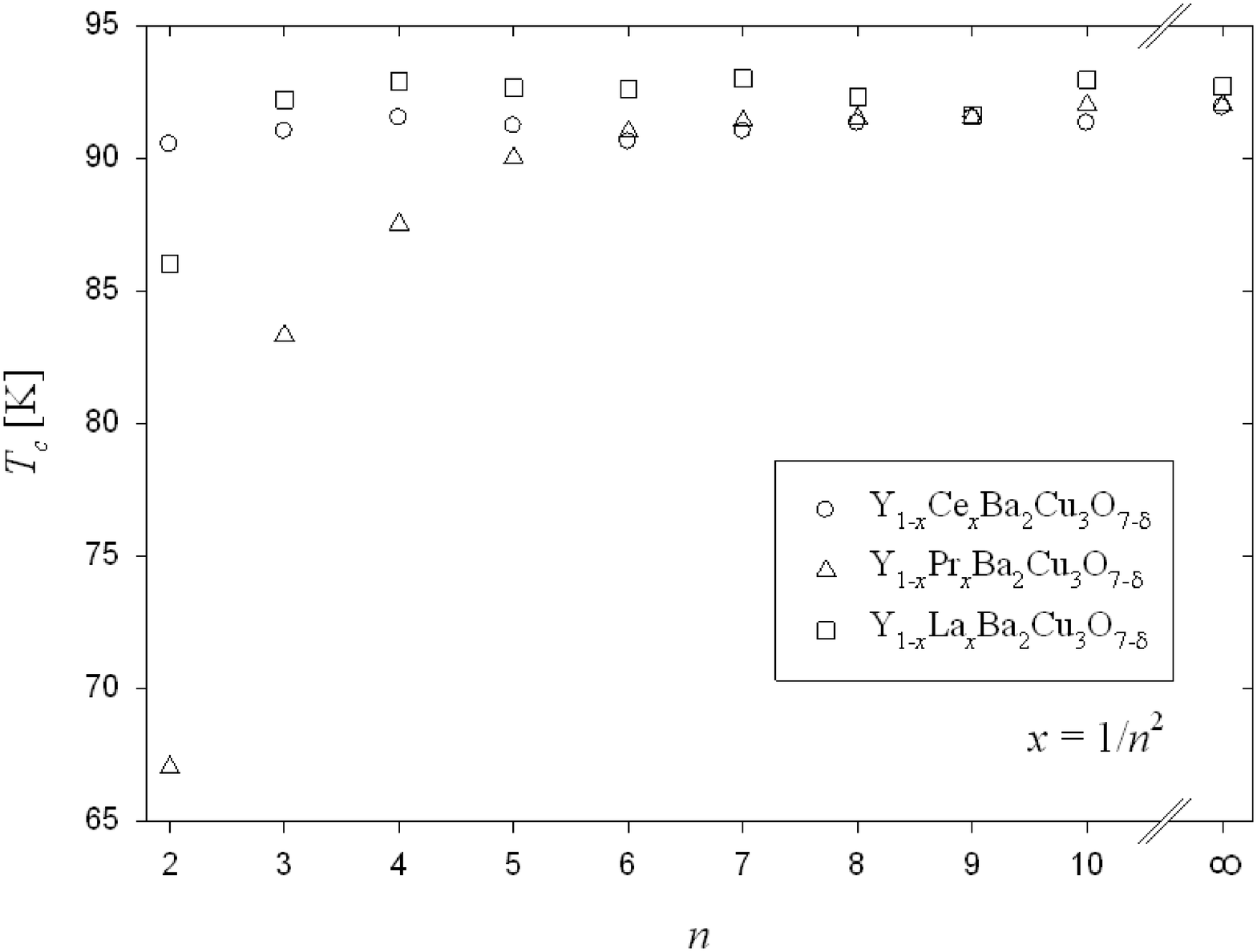}}
\caption{Intragrain critical temperature $T_{c}$ of
Y$_{1-x}$RE$_{x}$Ba$_{2}$Cu$_{3}$O$_{7-\delta}$.} \label{fig2}
\end{figure}

\begin{figure}[htbp]
\centerline{\includegraphics[width=6.04in,height=4.69in]{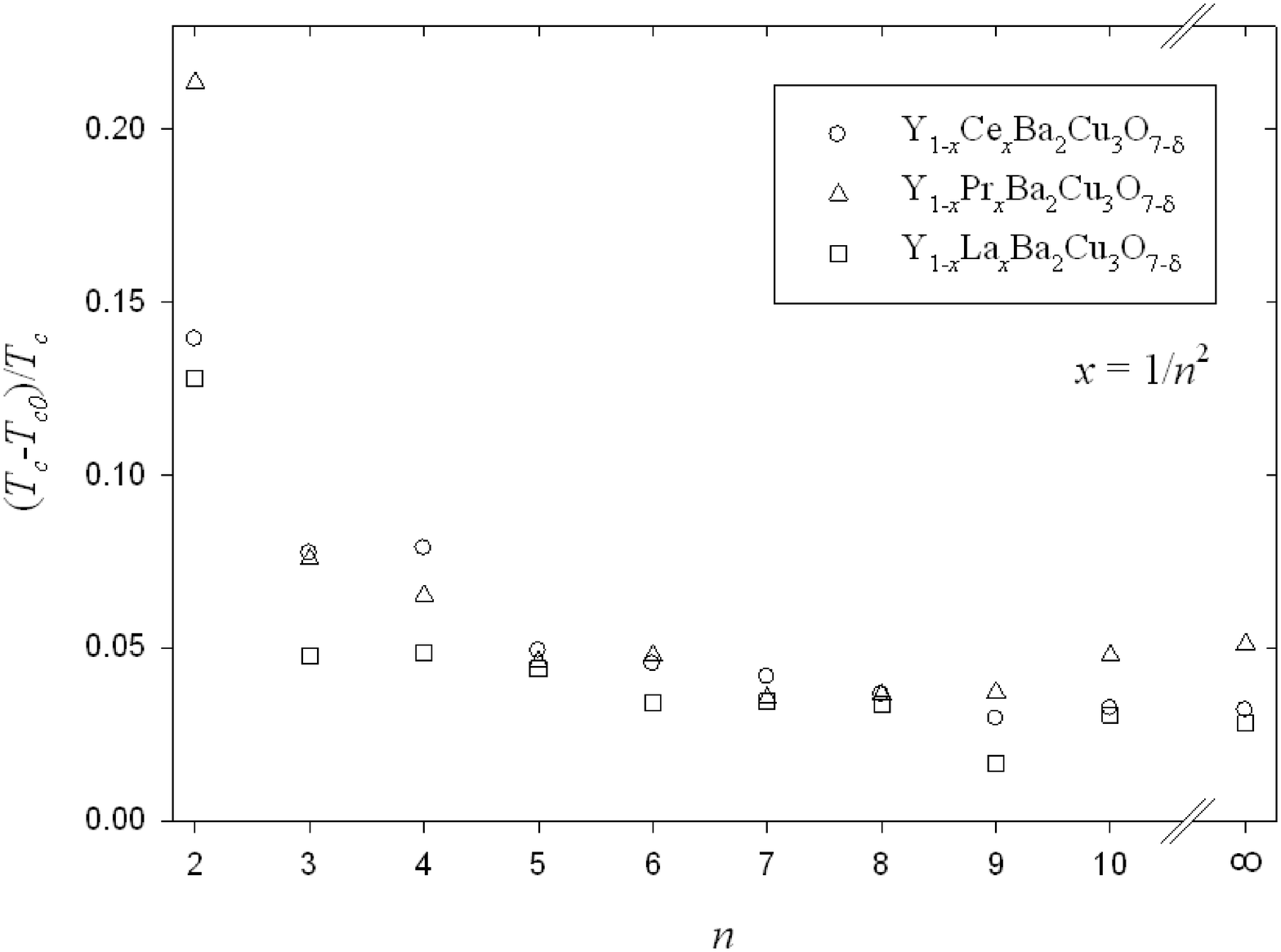}}
\caption{Normalized width of superconducting transition
($T_{c}-T_{c0})$/$T_{c}$ for
Y$_{1-x}$RE$_{x}$Ba$_{2}$Cu$_{3}$O$_{7-\delta}$.} \label{fig3}
\end{figure}

\Fref{fig2} demonstrates that $T_{c}$ of
Y$_{1-x}$RE$_{x}$Ba$_{2}$Cu$_{3}$O$_{7 - \delta }$ depends weakly on
RE concentration for $x <$ 0.0625 ($n >$ 4). To reveal the influence
of RE impurities to the intergrain boundaries we compared the width
of superconducting transition ($T_{c}-T_{c0})$ of samples. These
were normalized to $T_{c}$ for a correct comparison of the samples
with different $T_{c}$. \Fref{fig3} plots the normalized width of
superconducting transition ($T_{c}-T_{c0})$/$T_{c}$ for the samples
as a function of $n$. The dispersion of ($T_{c}-T_{c0})$/$T_{c}$ is
$ \approx $ 4 {\%} for the compositions with $x  \le $ 0.04 ($n  \ge
$ 5). It follows that there is no remarkable influence of RE
concentration on intergrain currents at these concentrations.

\begin{figure}[htbp]
\centerline{\includegraphics[width=6.37in,height=4.65in]{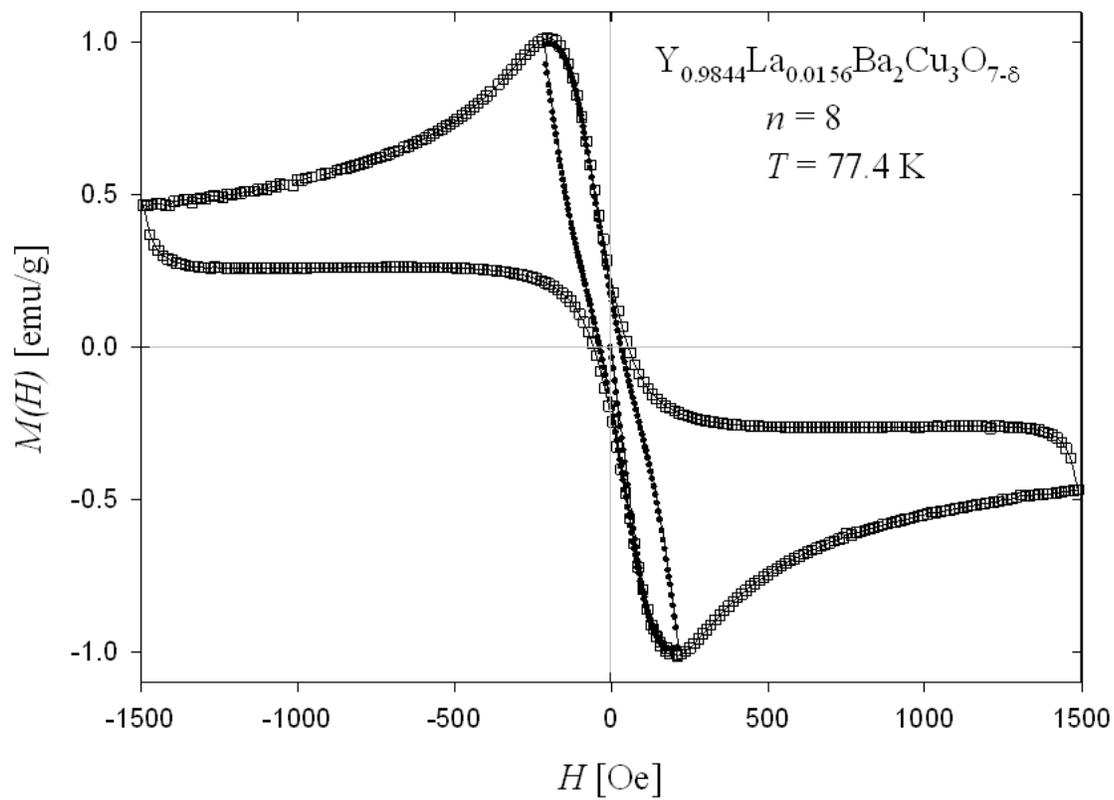}}
\caption{Magnetization of sample
Y$_{0.9844}$La$_{0.0156}$Ba$_{2}$Cu$_{3}$O$_{7 - \delta }$ vs.
magnetic field up to 200 Oe (black dots), 1500 Oe (squares).}
\label{fig4}
\end{figure}

Magnetization loops of samples are typical for polycrystalline
superconductors. Dependences $M(H)$ of sample
Y$_{1-x}$La$_{x}$Ba$_{2}$Cu$_{3}$O$_{7 - \delta }$ with $x$ = 0.0156
($n$ = 8) mesured up to different $H$ are presented in \Fref{fig4}.
There is the method \cite{mueller} separating the intergrain and
intragrain critical currents from curves $M(H)$ measured up to low
and high magnetic fields. However the width of loop $M(H)$ at zero
field is practically the same for curves measured up to 200 Oe and
1000 Oe. Also the observed asymmetry of loop $M(H)$ at high $H$ is a
sign of strong influence of the edge barriers \cite{shi,ain}. For
such case application of Bean model and the method \cite{mueller} is
incorrect \cite{yeshu}.

\begin{figure}[htbp]
\centerline{\includegraphics[width=4.20in,height=7.74in]{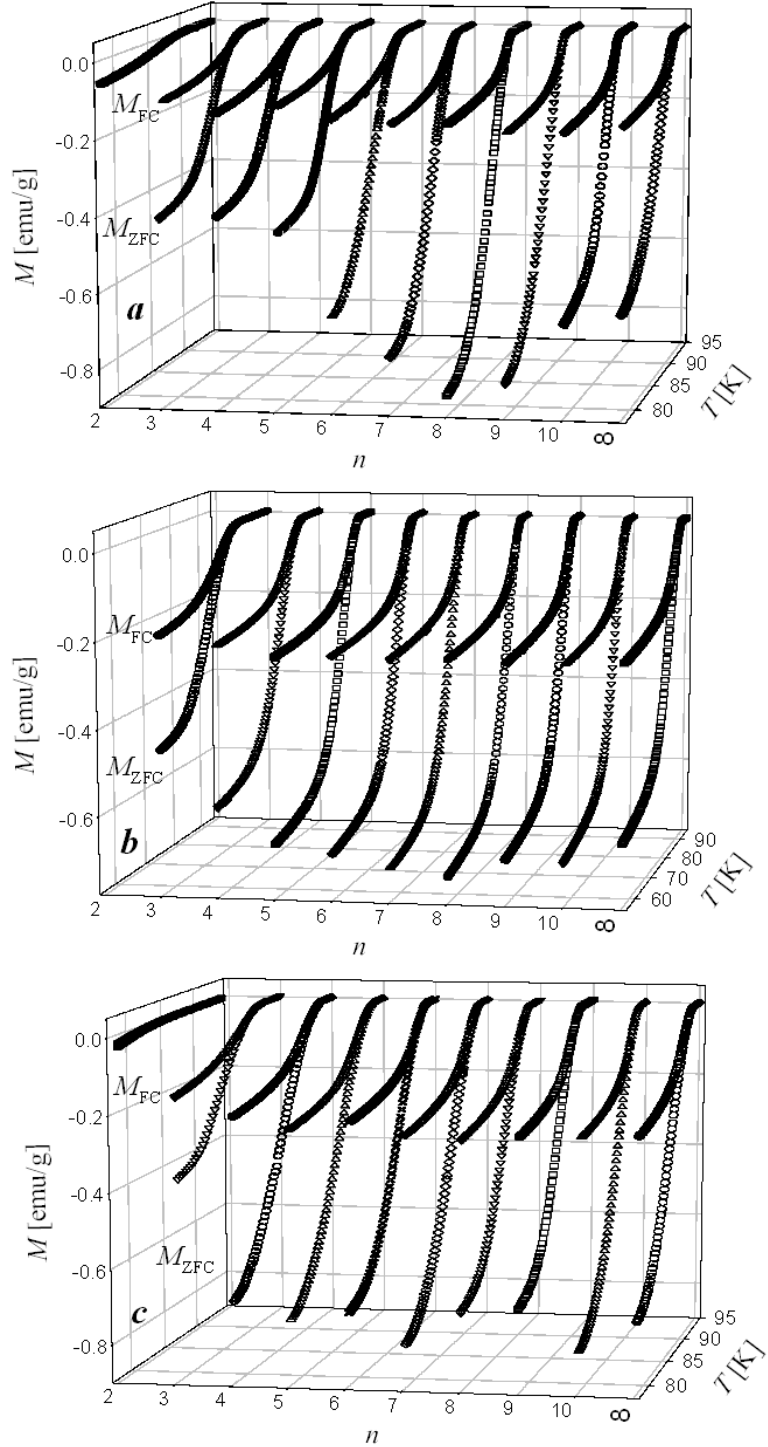}}
\caption{Temperature dependence of zero field cooled $M_{ZFC}(T)$
and field cooled $M_{FC}(T)$ magnetization of $a$)
Y$_{1-x}$Ce$_{x}$Ba$_{2}$Cu$_{3}$O$_{7-\delta}$, $b$)
Y$_{1-x}$Pr$_{x}$Ba$_{2}$Cu$_{3}$O$_{7-\delta}$, $c$)
Y$_{1-x}$La$_{x}$Ba$_{2}$Cu$_{3}$O$_{7-\delta}$.} \label{fig5}
\end{figure}

The temperature dependences of magnetizations $M_{FC}(T)$ and
$M_{ZFC}(T)$ of the samples at $H$ = 100 Oe are plotted in
\Fref{fig5} $a,b,c$. Data for
Y$_{0.75}$Pr$_{0.25}$Ba$_{2}$Cu$_{3}$O$_{7 - \delta }$ ($n$ = 2) are
lost unfortunately. It is clearly seen that the absolute values of
$M_{ZFC}(T)$ are sensitive to the concentration of RE. The sample
Y$_{1-x}$Pr$_{x}$Ba$_{2}$Cu$_{3}$O$_{7 - \delta }$ with $x$ = 0
($n=\infty )$ has somewhat smaller absolute values of $M_{ZFC}(T)$ at
any fixed $T$ than Y$_{1-x}$La$_{x}$Ba$_{2}$Cu$_{3}$O$_{7 - \delta
}$ and Y$_{1-x}$Ce$_{x}$Ba$_{2}$Cu$_{3}$O$_{7 - \delta }$ with $x$ =
0. The different oxygen content in the different series of samples
is possibly reason for this mismatch.

\begin{figure}[htbp]
\centerline{\includegraphics[width=6.46in,height=7.27in]{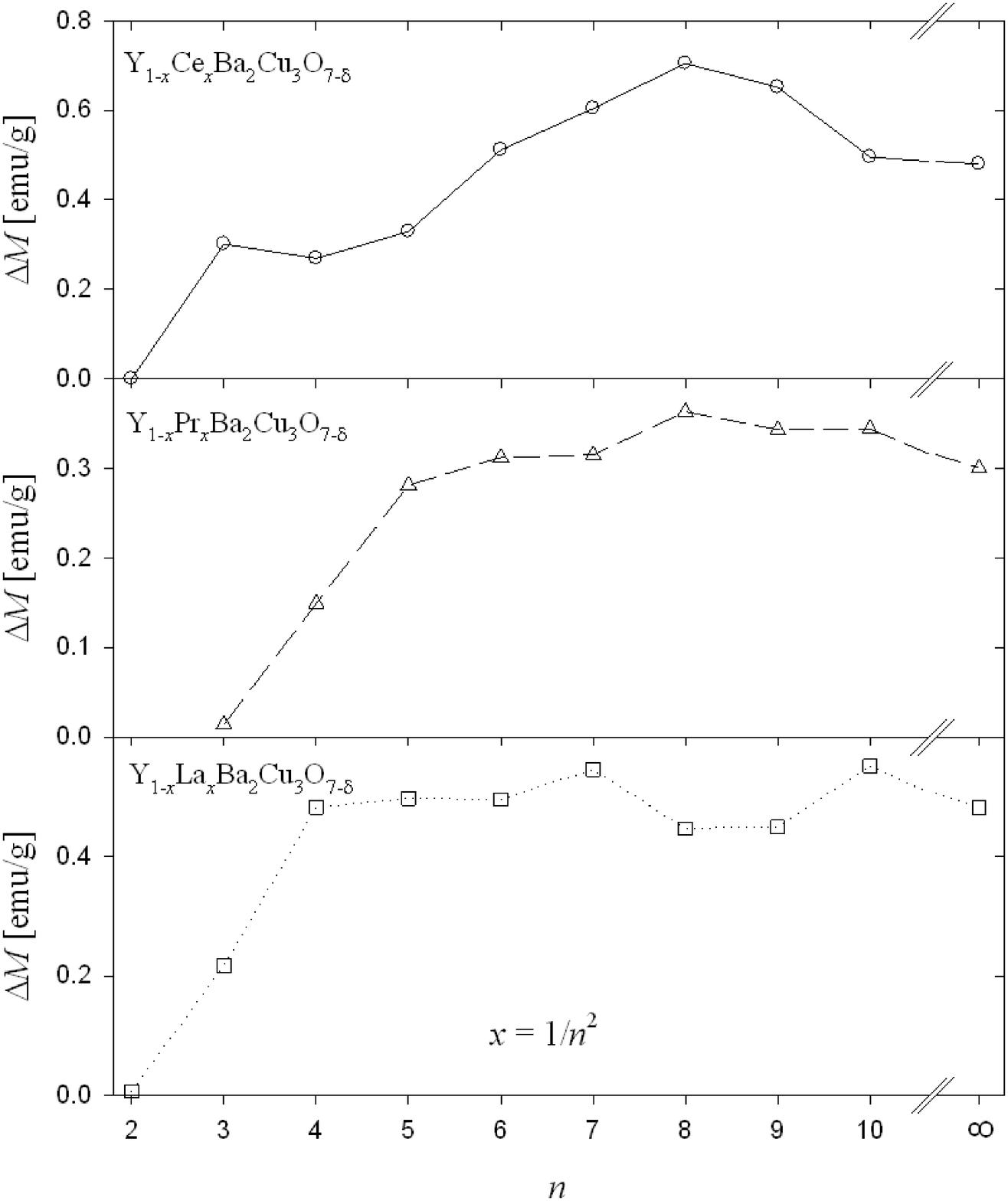}}
\caption{Difference of field cooled and zero field cooled
magnetization $\Delta M$ of
Y$_{1-x}$RE$_{x}$Ba$_{2}$Cu$_{3}$O$_{7-\delta}$ at 77.4 K.}
\label{fig6}
\end{figure}

The dependences $M_{FC}(T)$ and $M_{ZFC}(T)$ give information about
the pinning in samples. The difference between $M_{FC}(T)$ and
$M_{ZFC}(T)$ depends monotonically on the pinning energy in a type
II superconductor \cite{maloz}. Influence of the intergrain pinning
is small here due to the strong depinning in intergrain boundaries
\cite{jung}. Thus the value $\Delta M=M_{FC}-M_{ZFC}$ is concerned
with the intragrain pinning. \Fref{fig6} displays $\Delta M$ at 77.4
K as a function of $n$. The character of dependence $\Delta M(n)$
and the position of maximum do not change at other $T < T_{c}$. For
Y$_{1-x}$Ce$_{x}$Ba$_{2}$Cu$_{3}$O$_{7 - \delta }$ and
Y$_{1-x}$Pr$_{x}$Ba$_{2}$Cu$_{3}$O$_{7 - \delta }, \quad \Delta M$
has a maximal value at $n$ = 8 ($x$ = 0.0156). This position of
maximum corresponds to the average distance between the impurity
atoms $D$ = 8$a$. For YBCO $a$ = 0.382 nm, so $D$ = 3.06 nm. Such
value of $D$ is comparable to the coherence length in YBCO ($\xi
_{0} \sim $ 1.5 nm, $\xi$(77 K) $\sim $ 4 nm \cite{larb}). We
believe that the pinning is maximal then the defects spaced at the
average distances close to the diameter of Abrikosov vortex. Here
the pinning defects are the local distortions of electron density
formed by the RE ions with valence higher than 3+. The pinning
defects created by RE doping can also be associated with a formation
of oxygen vacancies in CuO$_2$ planes \cite{huht}.

The discrepancy between our result for maximal pinning concentration
and the result of article \cite{huht} ($x$ = 0.08) is possibly found
because the technique of synthesis used in investigation \cite{huht}
differs from ours. In the samples \cite{huht} the doping Pr
partially goes to intergrain boundaries modifying them. This leads
also to decreasing of the actual concentration of Pr in the
granules.

Y$_{1-x}$La$_{x}$Ba$_{2}$Cu$_{3}$O$_{7 - \delta }$ does not
demonstrate a clear maximum of $\Delta M$ for any $n$. The
distortions of crystal structure which arise from larger size of La
ions probably do not pin vortices. Therefore the doping by RE
elements with valence higher than 3+ is preferable for the high
pinning.

\section{Conclusions}

The resistance dependences on temperature of
Y$_{1-x}$RE$_{x}$Ba$_{2}$Cu$_{3}$O$_{7 - \delta }$ demonstrate that
the RE impurities do not modify the intergrain transport while $x <
0.04$. The remanent magnetization of the samples reveals that the
doping of the RE elements with valence higher than 3+ in small
amounts increases the intragrain pinning. The doping of YBCO by the
homovalent RE (La) does not increase the pinning. Concentration of
the heterovalent RE for the maximal pinning, x = 0.0156, is the same
for both tested elements (Ce and Pr). This concentration corresponds
to the average distance between the impurity atoms equal to 3.06 nm
($ \sim 2\xi _{0}$).

\section*{Acknowledgments}

We are thankful to A.D. Vasil'ev and M.S. Molokeev for XRD of
samples. The work is supported by Siberian Division of Russian
Academy of Sciences (complex integration project 3.4), Russian
Academy of Sciences (program ''Quantum Macrophysics'' project 3.4).

\newpage

\end{document}